\documentclass[aoas,nameyear,dvips]{arximspdf}
\usepackage{graphicx}
\doi{10.1214/10-AOAS356}
\volume{5}
\issue{4}
\pubyear{2011}
\firstpage{2278}
\lastpage{2299}


\begin{document}
\begin{frontmatter}

\title{Revisiting Guerry's data: Introducing spatial constraints in multivariate analysis}
\runtitle{Spatial multivariate analysis}

\begin{aug}
\author[A]{\fnms{St\'{e}phane} \snm{Dray}\ead[label=e1]{dray@biomserv.univ-lyon1.fr}\corref{}\ead[label=u1,url]{http://pbil.univ-lyon1.fr/members/dray}}
\and
\author[B]{\fnms{Thibaut} \snm{Jombart}\ead[label=e2]{t.jombart@imperial.ac.uk}}

\runauthor{S. Dray and T. Jombart}
\affiliation{Universit\'{e} Lyon 1 and Imperial College}
\address[A]{UMR 5558\\
  Laboratoire de Biom\'{e}trie et Biologie Evolutive\\
  Universit\'{e} de Lyon; universit\'{e} Lyon 1; CNRS\\
  43 boulevard du 11 novembre 1918\\
  Villeurbanne F-69622\\
  France\\
  \printead{e1}\\
  \printead{u1}} 
\address[B]{
  MRC Centre for Outbreak Analysis\\
  \quad\& Modelling\\
Department of Infectious\\
 \quad Disease Epidemiology\\
  Faculty of Medicine\\
  Imperial College London\\
  Norfolk Place\\
  London W2 1PG\\
  United Kingdom\\
  \printead{e2}}
\end{aug}

\received{\smonth{11} \syear{2009}}
\revised{\smonth{4} \syear{2010}}

\begin{abstract}
Standard multivariate analysis methods aim to identify and summarize
the main structures in large data sets containing the description of a
number of observations by several variables. In many cases, spatial
information is also available for each observation, so that a~map can
be associated to the multivariate data set. Two main objectives are
relevant in the analysis of spatial multivariate data: summarizing
covariation structures and identifying spatial patterns. In practice,
achieving both goals simultaneously is a statistical challenge, and a
range of methods have been developed that offer trade-offs between
these two objectives. In an applied context, this methodological
question has been and remains a major issue in community ecology, where
species assemblages (i.e., covariation between species abundances) are
often driven by spatial processes (and thus exhibit spatial patterns).

In this paper we review a variety of methods developed in community
ecology to investigate multivariate spatial patterns. We present
different ways of incorporating spatial constraints in multivariate
analysis and illustrate these different approaches using the famous
data set on moral statistics in France published by Andr\'{e}-Michel
Guerry in 1833. We discuss and compare the properties of these
different approaches both from a practical and theoretical viewpoint.

\end{abstract}


\end{frontmatter}

\section{Introduction}
A recent study [\citet{SD922}] revived Andr\'{e}-Michel Guerry's
(\citeyear{SD955}) \textit{Essai sur la Statistique Morale de la
France}. Guerry gathered data on crimes, suicide, literacy and other
``moral statistics'' for various d\'{e}partements (i.e., counties) in
France. He provided the first real social data analysis, using graphics
and maps to summarize this georeferenced multivariate data set. The
work of \citet{SD922} contained a historical part describing Guerry's
life and work in detail. In a second part, Friendly reanalyzed Guerry's
data using a variety of modern tools of multivariate and spatial
analysis. He considered two main approaches to analyzing a data set
involving both multivariate and geographical aspects: data-centric
(multivariate analysis) and map-centric (multivariate mapping)
displays. In the first approach, the multivariate structure is first
summarized using standard analysis methods [e.g., principal component analysis, \citet{SD308}] and visualization methods [e.g.,
biplot, \citet{SD203}]. The geographic information is only added a
posteriori to the graphs, using colors or other visual attributes.
This approach thus favors the display of multivariate structures over
spatial patterns. On the other hand, multivariate mapping (i.e., the
representation of several variables on a single map using multivariate
graphs) emphasizes the geographical context but fails to provide a
relevant summary of the covariations between the variables. Moreover,
multivariate mapping raises several technical issues such as the lack
of readability of multivariate symbols (e.g., Chernoff faces), which
can only be used to represent a few  variables and are sometimes
difficult for nonspecialists to interpret. \citet{SD922} stated that
Guerry's \textit{questions, methods and data still present challenges
for multivariate and spatial visualization today}. While he
acknowledged progress in both exploratory spatial data analysis and
multivariate methods, he also suggested that \textit{the integration of
these data-centric and map-centric visualization and analysis is still
incomplete}. He concluded his paper with a motivating question:
\textit{Who will rise to Guerry's challenge?}.

This challenge has been one of the major methodological concerns in
community ecology (and in other disciplines, e.g., public health) over
the last few decades. Community ecology is a subdiscipline of ecology
that aims to understand the organization and causes of species
associations. As community data are essentially multivariate (many
species, many sites, many environmental factors and complex
spatio-temporal sampling designs), questions about the structure and
drivers of ecological communities have traditionally been addressed
through multivariate analyses [\citet{SD406}]. Hence, it has been and
remains a very fertile field for the development and the application of
multivariate techniques. One of the most active research goals in
ecology today is to understand the relative importance of processes
that determine the spatial organization of biodiversity at multiple
scales [\citet{SD399}]. As a consequence, the last decade has seen
efforts in the methodological domain to render the multivariate
analysis of community data more spatially explicit or, conversely, to
generalize analyses of spatial distributions to handle the covariation
of many species. These methods allow us to identify the main spatial
patterns by considering simultaneously both multivariate and
geographical aspects of the data. They thus represent a first step
toward the integration of data-centric and map-centric visualizations
into a single method.

In this paper we take up Friendly's challenge by demonstrating how
several spatially-explicit multivariate methods developed initially in
the context of community ecology could also be of benefit to other
fields. We present different ways of incorporating the spatial
information into multivariate analysis, using the duality diagram
framework [\citet{SD182}] to describe the mathematical properties of
these methods. We illustrate these different methodological
alternatives by reanalyzing Guerry's data.

\section{Standard approaches}
$\!\!$We use the data set compiled by Michael Friend\-ly and available at
\url{http://www.math.yorku.ca/SCS/Gallery/guerry/}.\break This data set has
been recently analyzed by \citet{SD991} to illustrate a new interactive
visualization tool and is now distributed in the form of an R package
[see \citet{DJ2010} for details]. We consider six key quantitative
variables (Table \ref{tab:tab1}) for each of the 85 d\'{e}partements of
France in 1830 (Corsica, an island and often an outlier, was excluded).
In this section we focus on classical approaches that consider either
the multivariate or the spatial aspect of the data. In the next
sections we will present methods that consider both aspects
simultaneously.

\begin{table}
\tablewidth=302pt
  \caption{Variable names, labels and descriptions.
  Note that four variables have been recorded in the form of ``Population per...'' so that
   low values correspond to high rates, whereas high values correspond to low rates.
   Hence, for all of the variables, more (larger numbers) is ``morally'' better}\label{tab:tab1}
   \begin{tabular*}{302pt}{@{\extracolsep{\fill}}ll@{}}
    \hline
    \textbf{Label} & \textbf{Description}\\
    \hline
    Crime\_pers & Population per crime against persons\\
    Crime\_prop & Population per crime against property\\
    Literacy & Percent of military conscripts who can read and write\\
    Donations & Donations to the poor\\
    Infants & Population per illegitimate birth\\
    Suicides & Population per suicide\\
    \hline
  \end{tabular*}
\end{table}

\subsection{Multivariate analysis}
Multivariate analysis allows us to identify and summarize the primary
underlying structures in large data sets by removing any redundancy in
the data. It aims to construct a low-dimensional space (e.g., 2 or~3
dimensions) that retains most of the original variability of the data.
The classical output consists of graphical summaries of observations
and variables that are interpreted for the first few dimensions.

\subsubsection{The duality diagram theory}
Multivariate data are usually recorded in a matrix $\mathbf{X}$ with
$n$ rows (observations) and $p$ columns (variables). The duality
diagram is a mathematical framework that defines a multivariate
analysis setup using a set of three matrices. We can consider the
(possibly transformed) data matrix \textbf{X} ($n \times p$) as a part
of a statistical triplet $ ( \mathbf{X}, \mathbf{Q}, \mathbf{D}
 )$, where \textbf{Q} ($p \times p$) and \textbf{D} ($n \times
n$) are usually symmetric positive definite matrices used as metrics
[i.e., \textbf{Q} provides a metric for the variables (columns of
\textbf{X}) and \textbf{D} provides a metric for the observations (rows
of \textbf{X})]. This unifying mathematical framework encompasses very
general properties, which will be described, to the analysis of a
triplet. For more details, the reader should consult \citet{SD182},
\citet{SD833} or \citet{SD839}. The mathematical properties of each
particular method (corresponding to a particular choice of matrices
$\mathbf{X}, \mathbf{Q}$ and $\mathbf{D}$) can then be derived
from the general properties of the diagram. Note that the analysis of
the duality diagram associated to the triplet $ ( \mathbf{X},
\mathbf{Q}, \mathbf{D}  )$ is equivalent to the generalized
singular value decomposition [GSVD, e.g., \citet{SD260}, Appendix A] of
\textbf{X} with the metrics \textbf{Q} and \textbf{D}.

The analysis of the diagram consists of the eigen-decomposition of the
operators $\mathbf{X} \mathbf{Q} \mathbf{X}^\mathrm{T} \mathbf{D} $ or
$\mathbf{X}^\mathrm{T} \mathbf{D} \mathbf{X} \mathbf{Q} $. These two
eigen-decompositions are related to each other (\textit{dual}) and have
the same eigenvalues. Thus, we have
\begin{eqnarray*}
\mathbf{X} \mathbf{Q} \mathbf{X}^\mathrm{T} \mathbf{D} \mathbf{K} &=& \mathbf{K}
\bolds{\Lambda}_{[r]},\\
\mathbf{X}^\mathrm{T}\mathbf{D}\mathbf{X} \mathbf{Q} \mathbf{A} &= & \mathbf{A} \bolds{\Lambda}_{[r]}.
\end{eqnarray*}

$r$ is called the rank of the diagram, and the nonzero eigenvalues
$\lambda_1>\lambda_2>\cdots>\lambda_r>0$ are stored in the diagonal
matrix $\bolds{\Lambda}_{[r]}$.

$\mathbf{K}= [ \mathbf{k}^1,\ldots,\mathbf{k}^r  ]$ is a $n
\times r$ matrix containing the $r$ nonzero associated eigenvectors (in
columns).  These vectors are $\mathbf{D}$-orthonormalized (i.e.,\break
\mbox{$\mathbf{K}^\mathrm{T}\mathbf{D}\mathbf{K}=\mathbf{I}_r$}) and are
usually called the \textit{principal components}.

$\mathbf{A}= [ \mathbf{a}^1,\ldots,\mathbf{a}^r  ]$ is a $p
\times r$ matrix containing the $r$ nonzero eigenvectors (in columns).
These vectors are $\mathbf{Q}$-orthonormalized (i.e.,
$\mathbf{A}^\mathrm{T}\mathbf{Q}\mathbf{A}=\mathbf{I}_r$) and are
usually called the \textit{principal axes}.

The row scores $\mathbf{R}=\mathbf{X}\mathbf{Q}\mathbf{A}$ are obtained
by projection of the observations (rows of $\mathbf{X}$) onto the
principal axes. The vectors $\mathbf{a}^1,
\mathbf{a}^2,\dots,\mathbf{a}^r$ successively maximize, under the
$\mathbf{Q}$-orthogonality constraint, the following quadratic form:
\begin{equation}\label{eq0a}
Q(\mathbf{a})=\mathbf{a}^\mathrm{T}\mathbf{Q}^\mathrm{T}\mathbf{X}^\mathrm{T}\mathbf{DXQa}.
\end{equation}
If $\mathbf{D}$ defines a scalar product, then we have
$Q(\mathbf{a})=\Vert\mathbf{X}\mathbf{Q}\mathbf{a}\Vert^2_\mathbf{D}$.

The column scores
$\mathbf{C}=\mathbf{X}^\mathrm{T}\mathbf{D}\mathbf{K}$ are obtained by
projection of the variables (columns of $\mathbf{X}$) onto the
principal components. The vectors $\mathbf{k}^1,
\mathbf{k}^2,\ldots,\mathbf{k}^r$ successively maximize, under the
$\mathbf{D}$-orthogonality constraint, the following quadratic form:
\begin{equation}\label{eq0b}
S(\mathbf{k})=\mathbf{k}^\mathrm{T}\mathbf{D}^\mathrm{T}\mathbf{XQX}^\mathrm{T}\mathbf{Dk}.
\end{equation}
If $\mathbf{Q}$ defines a scalar product, then we have
$S(\mathbf{k})=\Vert\mathbf{X}^\mathrm{T}\mathbf{D}\mathbf{k}\Vert^2_\mathbf{Q}$.
Usually, the outputs (column and row scores) are only interpreted for
the first few axes (dimensions).
\subsubsection{Application to Guerry's data}
Here we consider $p=6$ variables measured for $n=85$ observations
(d\'{e}partements of France). As only quantitative variables have been
recorded, principal component analysis [PCA, \citet{SD308}] is well
adapted. Applying PCA to the correlation matrix where
$\mathbf{Q}=\mathbf{I}_p$, $\mathbf{D}=\frac{1}{n} \mathbf{I}_n$ and
$\mathbf{X}$ contains \textit{z}-scores, we obtain
$Q(\mathbf{a})=\Vert\mathbf{X}\mathbf{Q}\mathbf{a}\Vert^2_\mathbf{D}=\operatorname{var}(\mathbf{XQa})$
and
$S(\mathbf{k})=\Vert\mathbf{X}^\mathrm{T}\mathbf{D}\mathbf{k}\Vert^2_\mathbf{Q}=\sum_{j=1}^{p}
\operatorname{cor}^2(\mathbf{k},\mathbf{x}^j)$ from equations (\ref{eq0a}) and
(\ref{eq0b}). Hence, this PCA summarizes the data by maximizing
simultaneously the variance of the projection of the observations onto
the principal axes and the sum of the squared correlations between the
principal component and the variables.

\begin{figure}

\includegraphics{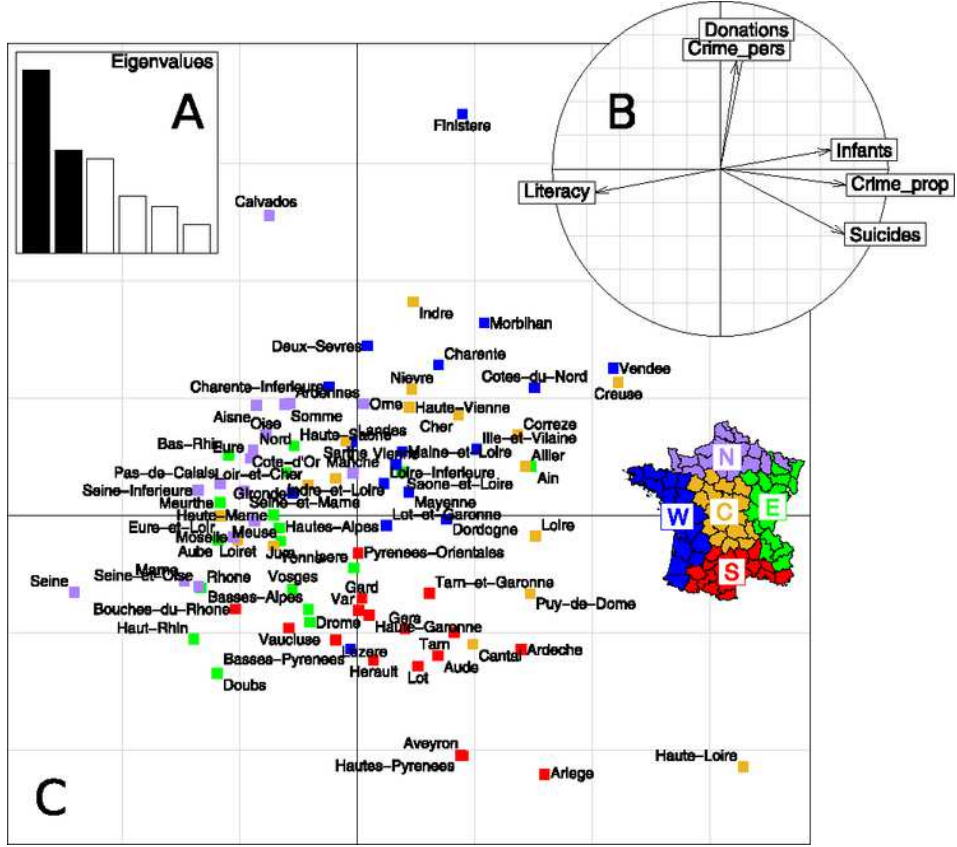}

\caption{Principal component analysis of Guerry's data.
\textup{(A)} Barplot of eigenvalues. \textup{(B)} Correlation between variables and
principal components. \textup{(C)} Projections of d\'{e}partements on principal
axes. The color of each square corresponds to a region of
France.}\label{fig:fig1}
\vspace*{3pt}
\end{figure}

For didactic purposes, following \citet{SD922}, we interpret two
dimensions, while the barplot of eigenvalues (Figure \ref{fig:fig1}A)
would rather suggest a~1-D or a 3-D solution. The first two PCA
dimensions account for 35.7\% and 20\%, respectively, of the total
variance. The correlations between variables and principal components
are represented on the correlation circle in Figure \ref{fig:fig1}B. As
we have excluded Corsica (an outlier) in the present paper, the results
are slightly different from those reported in \citet{SD922}. The first
axis is negatively correlated to literacy and positively correlated to
property crime, suicides and illegitimate births. The second axis is
aligned mainly with personal crime and donations to the poor. As we are
also interested in spatial patterns, we have added geographical
information in the form of color symbols on the factorial map of
d\'{e}partements (Figure \ref{fig:fig1}C). Each color corresponds to one
of five regions of France. The results are quite difficult to
interpret, but some general patterns can be reported. For the first
axis, the North and East are characterized by negative scores,
corresponding to high levels of literacy and high numbers of suicides,
crimes against property and illegitimate births. The second axis mainly
contrasts the West (high donations to the the poor and low levels of
crime against persons) to the South.

\subsection{Spatial autocorrelation}
Exploratory spatial data analysis (ESDA) is a subset of exploratory
data analysis [EDA, \citet{SD934}] that focuses on detecting spatial
patterns in data [\citet{SD919}]. In this context, spatial
autocorrelation statistics, such as \citet{SD455}'s Coefficient (MC)
and the \citet{SD223} Ratio, aim to measure and analyze the degree of
dependency among observations in a geographical context [\citet{SD577}].

\subsubsection{The spatial weighting matrix}
The first step of spatial autocorrelation analysis is to define a  $n
\times n$ spatial weighting matrix, usually denoted~\textbf{W}. This
matrix is a mathematical representation of the geographical layout of
the region under study [\citet{SD961}]. The spatial weights reflect
a priori the absence ($w_{ij} =0$), presence or intensity
($w_{ij} >0$) of the spatial relationships between the locations
concerned. Spatial weighting matrices can be usefully represented as
graphs (neighborhood graphs), where nodes correspond to spatial units
(d\'{e}partements) and edges to nonnull spatial weights.\looseness=1

\begin{figure}

\includegraphics{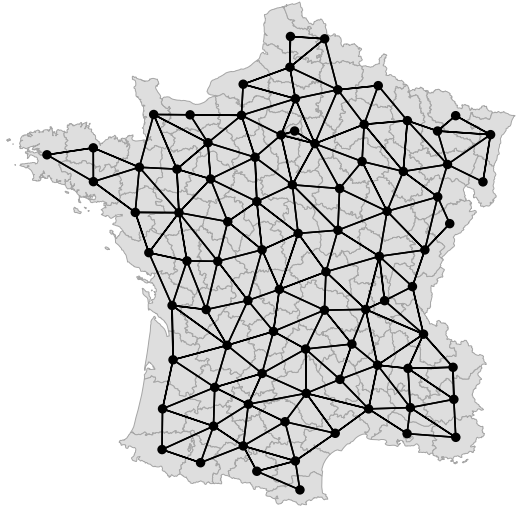}

\caption{Neighborhood relationships between d\'{e}partements of France.}\label{fig:fig2}
\end{figure}

The simplest neighborhood specification is a connectivity matrix
\textbf{C}, in which $c_{ij} =1$ if spatial units $i$ and $j$ are
neighbors and $c_{ij} =0$ otherwise. More sophisticated definitions
[\citet{SD224}; \citet{SD163}] are able to take into account the distances between
the spatial units or the length of the common boundary between the
regions for areal data. In the case of Guerry's data, we simply defined
a binary neighborhood where two d\'{e}partements $i$ and $j$ are
considered as neighbors ($c_{ij} =1$) if they share a common border
(Figure~\ref{fig:fig2}).

The connectivity matrix \textbf{C} is usually scaled to obtain a
spatial weighting matrix \textbf{W}, most often with zero diagonal. The
row-sum standardization (elements sum to 1 in each row) is generally
preferred; it is obtained by
\[
w_{ij}=\frac{c_{ij}}{\sum_{j=1}^n{c_{ij}}}.
\]

Alternative standardizations are discussed in \citet{SD664}.

\subsubsection{Moran's coefficient}
Once the spatial weights have been defined, the spatial autocorrelation
statistics can then be computed. Let us consider the $n$-by-1 vector
$\mathbf{x}=[ {x_1 \cdots x_n } ]^\mathrm{T}$ containing
measurements of a quantitative variable for $n$ spatial units. The
usual formulation for Moran's coefficient of spatial autocorrelation
[\citet{SD577}; \citet{SD455}] is
\begin{equation}\label{eq1}
\hspace*{16pt}\operatorname{MC}(\mathbf{x})=\frac{n\sum_{( 2 )}
{w_{ij} (x_i -\bar {x})(x_j -\bar {x})} }{\sum_{( 2
)} {w_{ij} } \sum_{i=1}^n {(x_i -\bar {x})^2} }\qquad\mbox{where }\sum_{( 2 )} =\sum_{i=1}^n
{\sum_{j=1}^n }\mbox{ with }i\ne j.
\end{equation}

MC can be rewritten using matrix notation:
\begin{equation}\label{eq2}
\operatorname{MC}(\mathbf{x})=\frac{n}{\mathbf{1}^\mathrm{T}\mathbf{W1}}\frac{\mathbf{z}^\mathrm{T}{\mathbf{Wz}}}{\mathbf{z}^\mathrm{T}\mathbf{z}},
\end{equation}
where $\mathbf{z}= (
\mathbf{I}_n-\mathbf{1}_n\mathbf{1}_n^\mathrm{T} /n  )\mathbf{x}$
is the vector of centered values ($z_i=x_i-\bar{x}$) and~$\mathbf{1}_n$
is a vector of ones (of length $n$).

The numerator of MC corresponds to the covariation between contiguous
observations. This covariation is standardized by the denominator,
which measures the variance among the\vadjust{\goodbreak} observations. The significance of
the observed value of MC can be tested by a Monte Carlo procedure, in
which locations are permuted to obtain a distribution of MC under the
null hypothesis of random distribution. An observed value of MC that is
greater than that expected at random indicates the clustering of
similar values across space (positive spatial autocorrelation), while a
significant negative value of MC indicates that neighboring values are
more dissimilar than expected by chance (negative spatial
autocorrelation).

We computed MC for Guerry's data set using the row-standardized
definition of the spatial weighting matrix associated with the
neighborhood graph presented in Figure \ref{fig:fig2}. A positive and
significant autocorrelation is identified for each of the six variables
(Table \ref{tab:tab2}). Thus, the values of literacy are the most
covariant in adjacent departments, while illegitimate births (Infants)
covary least.

\begin{table}
\tablewidth=200pt \caption{Values of Moran's coefficient for the six
variables. P-values obtained by a randomization testing procedure (999
permutations) are given in parentheses}\label{tab:tab2}
\begin{tabular*}{200pt}{@{\extracolsep{\fill}}lc@{}}
\hline
& \textbf{MC}\\
\hline
Crime\_pers & 0.411 (0.001)\\
Crime\_prop & 0.264 (0.001)\\
Literacy & 0.718 (0.001)\\
Donations & 0.353 (0.001)\\
Infants & 0.229 (0.001)\\
Suicides & 0.402 (0.001)\\
\hline
\end{tabular*}
\end{table}

\subsubsection{Moran scatterplot}
If the spatial weighting matrix is row-standar\-dized, we can define the
lag vector $\mathbf{\tilde{z}} = \mathbf{Wz}$ (i.e., $\tilde{z}_i =
\sum_{j=1}^n{w_{ij}x_j}$) composed of the weighted (by the spatial
weighting matrix) averages of the neighboring values. Equation
(\ref{eq2}) can then be rewritten as
\begin{equation}
\label{eq3}
\operatorname{MC}(\mathbf{x})=\frac{\mathbf{z}^\mathrm{T}{\mathbf{\tilde{z}}}}{\mathbf{z}^\mathrm{T}\mathbf{z}},
\end{equation}
since in this case $\mathbf{1}^\mathrm{T}\mathbf{W1}=n$. Equation
(\ref{eq3}) shows clearly that MC measures the autocorrelation by giving
an indication of the intensity of the linear association between the
vector of observed values $\mathbf{z}$ and the vector of weighted
averages of neighboring values $\mathbf{\tilde{z}}$. \citet{SD566}
proposed to visualize MC in the form of a bivariate scatterplot of
$\mathbf{\tilde{z}}$ against $\mathbf{z}$. A linear regression can be
added to this \textit{Moran scatterplot}, with slope equal to MC. The
Moran scatterplot is a very nice graphical tool to evaluate and
represent the degree of spatial autocorrelation, the presence of
outliers or local pockets of nonstationarity [\citet{SD565}].

\begin{figure}

\includegraphics{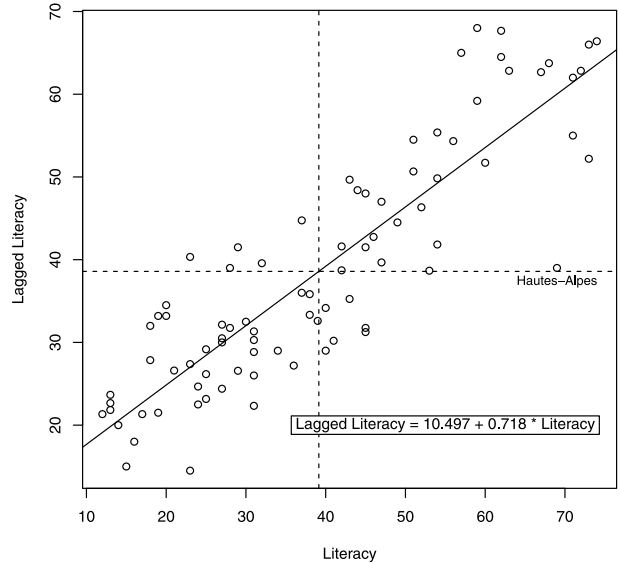}

  \caption{Moran scatterplot for Literacy. Dotted lines corresponds to means.}
  \label{fig:fig3}
\end{figure}

Considering the Literacy variable of Guerry's data, the Moran
scatterplot (Figure \ref{fig:fig3}) clearly shows strong
autocorrelation. It also shows that the Hautes-Alpes d\'{e}partement
has a slightly outlying position characterized by a high value of
Literacy compared to its neighbors. This d\'{e}partement can be
considered as a leverage point that drags down the assessment of the
link between Literacy and spatial-lagged literacy (i.e., MC). This is
confirmed by different diagnostic tools [DFFITS, Cook's D, e.g.,
\citet{SD990}] adapted to the linear model.

\subsection{Toward an integration of multivariate and geographical aspects}
\begin{figure}

\includegraphics{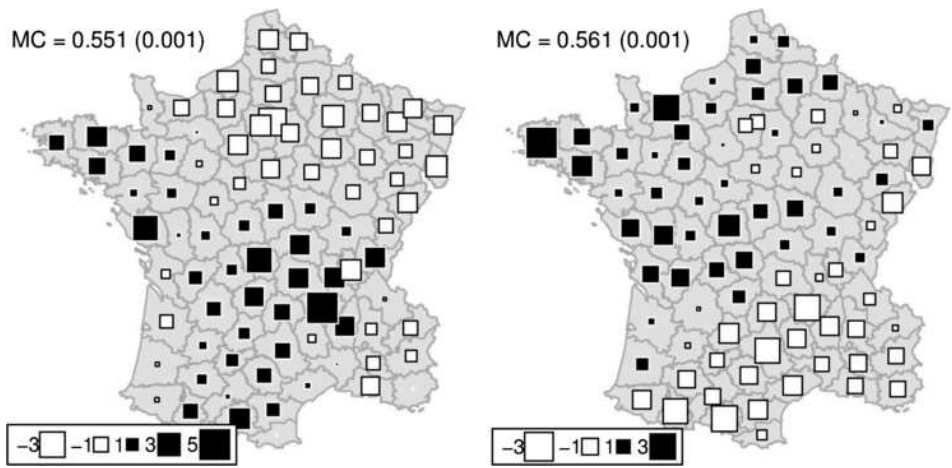}
 \caption{Principal component analysis of Guerry's
data. Map of d\'{e}partements' scores for the first (left) and
second (right) PCA axes. Values of Moran's coefficient and
associated P-values obtained by a randomization testing procedure (999
permutations) are given.}\label{fig:fig4}
\vspace*{-3pt}
\end{figure}

The integration of multivariate and spatial information has a long
history in ecology. The simplest approach considered a two-step
procedure where the data are first summarized with multivariate
analysis such as PCA. In a~second step, univariate spatial statistics
or mapping techniques are applied to PCA scores for each axis
separately. \citet{SD241} was the first to apply multivariate analysis
in ecology, and he integrated spatial information a~posteriori
by mapping PCA scores onto the geographical space using contour lines.
One can also test for the presence of spatial autocorrelation for the
first few scores of the analysis, with univariate autocorrelation
statistics such as MC. For instance, we mapped scores of the
d\'{e}partements for the first two axes of the PCA of Guerry's data
(Figure \ref{fig:fig4}). Even if PCA maximizes only the variance of these
scores, there is also a clear spatial structure, as the scores are
highly autocorrelated. The map for the first axis corresponds closely
to the split between \textit{la France \'{e}clair\'{e}e} (North-East
characterized by an higher level of Literacy) and \textit{la France
obscure}.

It is very simple to carry out this two-step approach but it has the
major disadvantage of being indirect, as it considers the spatial
pattern only after summarizing the main structures of the multivariate
data set. \citet{SD567} proposed a more direct approach by extending
the Moran scatterplot to the bivariate case. If we consider two
centered variables $\mathbf{z}_1$ and $\mathbf{z}_2$, the bivariate
Moran scatterplot represents $\mathbf{\tilde{z}}_2=\mathbf{Wz}_2$ on
the vertical axis and $\mathbf{z}_1$ on the horizontal axis. In a case
with more than two variables, one can produce bivariate Moran
scatterplots for all combinations of pairs of variables. However, this
approach becomes difficult to use when the number of variables
increases. In the next section we present several approaches that go
one step further by considering the identification of spatial
structures and the dimensionality reduction simultaneously.

\section{Spatial multivariate analysis}
Over the last two decades, several approaches have been developed to
consider both geographical and multivariate information simultaneously.
The multivariate aspect is usually treated by techniques of
dimensionality reduction similar to PCA. On the other hand, several
alternatives have been proposed to integrate the spatial information.
We review various alternatives in the following sections.
\subsection{Spatial partition}
One alternative is to consider a spatial partition of the study area.
In this case, the spatial information is coded as a catego\-rical
variable, and each category corresponds to a region of the whole study
area.\vadjust{\goodbreak} This partitioning can be inherent to the data set (e.g.,
administrative units) or can be constructed using geographic
information systems [e.g., grids of varying cell size in
\citet{SD165}]. For instance, Guerry's data contained a partition of
France into 5 regions (Figure~\ref{fig:fig1}).

In this context, searching for multivariate spatial structures would
lead us to look for a low-dimensional view that maximizes the
difference between the regions. To this end, \citet{SD922} used
discriminant analysis, a widely-used method providing linear
combinations of variables that maximize the separation between groups
as measured by an univariate $F$ statistic. However, this method
suffers from some limitations: it requires the number of variables to
be smaller than the number of observations, and it is impaired by
collinearity among variables. Here we used an alternative and lesser
known approach, the between-class analysis [BCA, \citet{SD148}], to
investigate differences between regions. Unlike discriminant analysis,
BCA maximizes the variance between groups (without accounting for the
variance within groups) and is not subject to the restrictions applying
to the former method.

BCA associates a triplet $ ( \mathbf{X}, \mathbf{Q}, \mathbf{D}
 )$ to a $n \times g$ matrix $\mathbf{Y}$ of dummy variables
indicating group membership. Let \textbf{A} be the $g \times p$ matrix
of group means for the \textit{p} variables\vadjust{\goodbreak} and $\mathbf{D_Y}$ be the
$g \times g$ diagonal matrix of group weights derived from the matrix
\textbf{D} of observation weights. By definition, we have
$\mathbf{A}=(\mathbf{Y}^\mathrm{T}\mathbf{DY})^{-1}\mathbf{Y}^\mathrm{T}\mathbf{DX}$
and $\mathbf{D_Y}=(\mathbf{Y}^\mathrm{T}\mathbf{DY})$. BCA corresponds
to the analysis of $ ( \mathbf{A}, \mathbf{Q}, \mathbf{D_Y}
)$ and diagonalizes the between-groups covariance matrix
$\mathbf{A}^\mathrm{T}\mathbf{D_YAQ}$.

\begin{figure}

\includegraphics{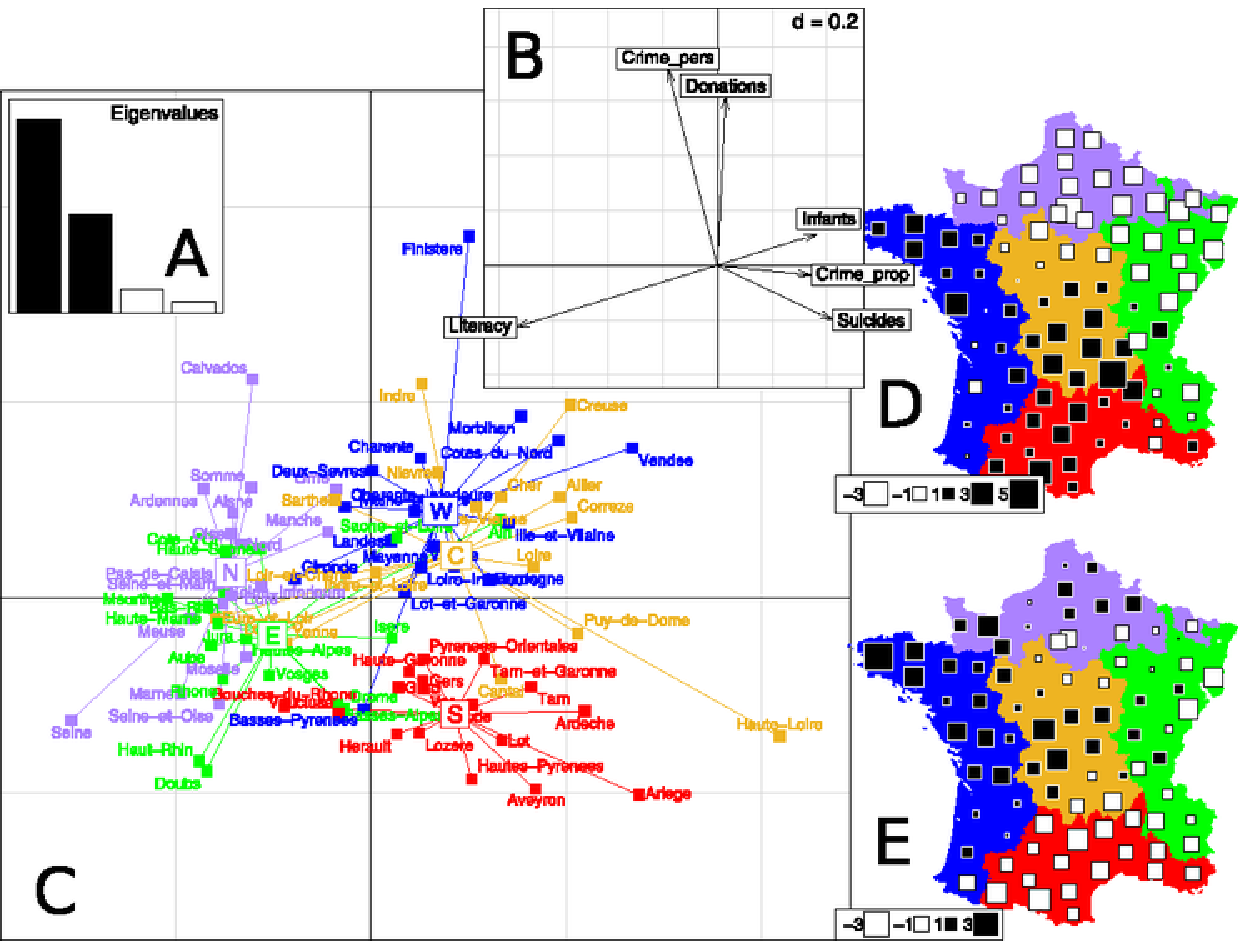}

\caption{Between-class analysis of Guerry's data.
\textup{(A)} Barplot of eigenvalues. \textup{(B)} Coefficients of
variables. \textup{(C)} Projections of d\'{e}partements on the BCA
axes. Map of d\'{e}partements scores for the first \textup{(D)} and
second \textup{(E)} axes. The different colors correspond to regions of
France.}\label{fig:fig5}\vspace*{-3pt}
\end{figure}

Here, 28.8\% of the total variance (sum of eigenvalues of PCA)
corresponds to the between-regions variance (sum of the eigenvalues of
BCA). The barplot of eigenvalues indicates that two axes should be
interpreted (Figure \ref{fig:fig5}A). The first two BCA dimensions
account for 59\% and 30.2\%, respectively, of the between-regions
variance. The coefficients used to construct the linear combinations of
variables are represented on Figure \ref{fig:fig5}B. The first axis
opposed literacy to property crime, suicides and illegitimate births.
The second axis is mainly aligned with personal crime and donations to
the poor. The factorial map of d\'{e}partements (Figure
\ref{fig:fig5}C) and the maps of the scores (Figure \ref{fig:fig5}D, E)
show the spatial aspects. The results are very close to those obtained
by PCA: the first axis contrasted the North and the East (\textit{la
France \'{e}clair\'{e}e}) to the other regions, while the South is
separated from the other regions by the second axis. The high
variability of the region Center is also noticeable. In contrast, the
South is very homogeneous.

\subsection{Spatial explanatory variables}
Principal component analysis with respect to the instrumental variables
[PCAIV, \citet{SD540}], also known as redundancy analysis
[\citet{SD674}], is a direct extension of PCA and multiple regression
adapted to the case of multivariate response data. The analysis
associates a $n \times q$ matrix  \textbf{Z} of explanatory variables
to the triplet\vadjust{\goodbreak} $ ( \mathbf{X}, \mathbf{Q}, \mathbf{D}  )$ where the
matrix \textbf{X} contains the response variables. The
\textbf{D}-orthogonal projector
$\mathbf{P_Z}=\mathbf{Z}(\mathbf{Z}^\mathrm{T}\mathbf{DZ})^{-1}\mathbf{Z}^\mathrm{T}\mathbf{D}$
is first used in a multivariate regression step to compute a matrix of
predicted values $\mathbf{\hat{X}}=\mathbf{P_ZX}$. The second step of
PCAIV consists~of the PCA of this matrix of predicted values and
corresponds then to the analysis of the~triplet $ ( \mathbf{\hat{X}},
\mathbf{Q}, \mathbf{D}  )$. Whereas PCA maximizes the variance of the
projection of the observations onto the principal axes, PCAIV maximizes
the variance explained by~$\mathbf{Z}$.

PCAIV and related methods, such as canonical correspondence analysis
[\citet{SD642}], have been often used in community ecology to identify
spatial relationships. The spatial information is introduced in the
matrix \textbf{Z} under the form of spatial predictors and the analysis
maximized then the ``spatial variance'' (i.e., the variance explained
by spatial predictors).  Note that BCA can also be considered as a
particular case of PCAIV, where the explanatory variables are dummy
variables indicating group membership.

\subsubsection{Trend surface of geographic coordinates}
From the EDA point of view, the data exploration has been
conceptualized by \citet{SD934} in the quasi-mathematical form\vadjust{\goodbreak}
$\mathit{DATA}=\mathit{SMOOTH}+\mathit{ROUGH}$. Trend surface analysis
is the oldest procedure for separating large-scale structure
($\mathit{SMOOTH}$) from random variation ($\mathit{ROUGH}$). Student
(\citeyear{SD626}) proposed expressing observed values in time series as
a polynomial function of time, and mentioned that this could be done
for spatial data as well. \citet{SD59} extended this approach to the
spatial and multivariate case by introducing polynomial functions of
geographic coordinates as predictors in PCAIV. We call this approach
PCAIV-POLY in the rest of the paper. Usually, polynomials of degree 2
or 3 are used; spurious correlations between these spatial predictors
can be removed using an orthogonalization procedure to obtain
orthogonal polynomials.

\begin{figure}

\includegraphics{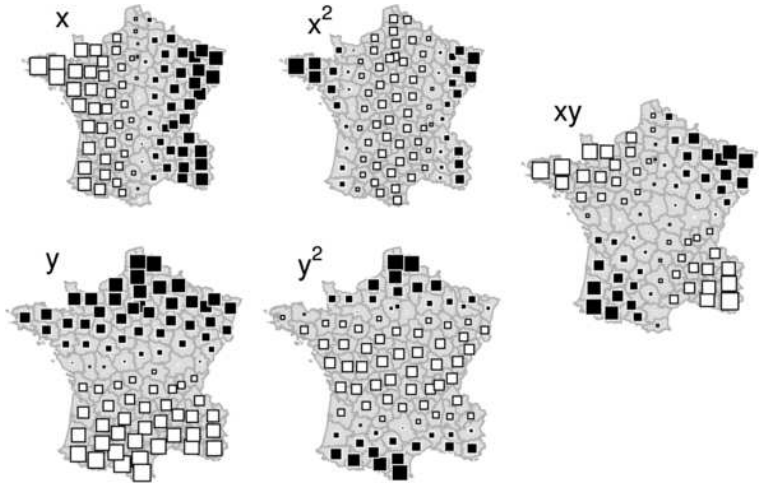}

\caption{Maps of the terms of a second-degree orthogonal polynomial.
Centroids of d\'{e}partements have been used as original
coordinates to construct the polynomial.}\label{fig:fig6}
\end{figure}

The centroids of d\'{e}partements of France were used to construct a
second-degree orthogonal polynomial (Figure \ref{fig:fig6}).

Here, 32.4\% of the total variance (sum of eigenvalues of PCA) is
explained by the second-degree polynomial (sum of eigenvalues of
PCAIV). The first two dimensions account for 51.4\% and 35.2\%,
respectively, of the explained variance. The outputs of PCAIV-POLY
(coefficients of variables, maps of d\'{e}partements scores, etc.) are
not presented, as they are very similar to those obtained by BCA.

\subsubsection{Moran's eigenvector maps}
An alternative way to build spatial predictors is by the
diagonalization of the spatial weighting matrix \textbf{W}.
\citet{SD135} have shown that the upper and lower bounds of MC for a
given spatial weighting matrix \textbf{W} are equal to  $\lambda_{\mathrm{max}}
(n/\mathbf{1}^\mathrm{T}\mathbf{W1})$ and $\lambda_{\mathrm{min}}
(n/\mathbf{1}^\mathrm{T}\mathbf{W1})$, where $\lambda_{\mathrm{max}}$ and
$\lambda_{\mathrm{min}}$ are the extreme eigenvalues of $\bolds{\Omega}=
\mathbf{HWH}$ where $\mathbf{H}= (
\mathbf{I}-\mathbf{11}^\mathrm{T} /n  )$ is a centering operator.
If a nonsymmetric spatial weighting matrix ${\mathbf{W}}^{*}$ has been
defined, the results can be generalized using
$\mathbf{W}=(\mathbf{W}^{*}+{\mathbf{W}^{*}}^\mathrm{T})/2$.

Moran's eigenvector maps [MEM, \citet{SD163}] are the $n-1$
eigenvectors of $\bolds{\Omega}$. They are orthogonal vectors with a
unit norm maximizing MC [\citet{SD264}]. MEM associated with high
positive (or negative) eigenvalues have high positive (or negative)
autocorrelation. MEM associated with eigenvalues with small absolute
values correspond to low spatial autocorrelation, and are not suitable
for defining spatial structures [\citet{SD163}]. Unlike polynomial
functions, MEM have the ability to capture various spatial structures
at multiple scales (coarse to fine scales). MEM have been used for
spatial filtering purposes [\citet{SD268}; \citet{SD226}] and introduced as
spatial predictors in linear models [Griffith (\citeyear{SD264}, \citeyear{SD266})], generalized
linear models [Griffith (\citeyear{SD267}, \citeyear{SD269})] and multivariate analysis
[\citet{SD163}; \citet{SD865}].

\begin{figure}

\includegraphics{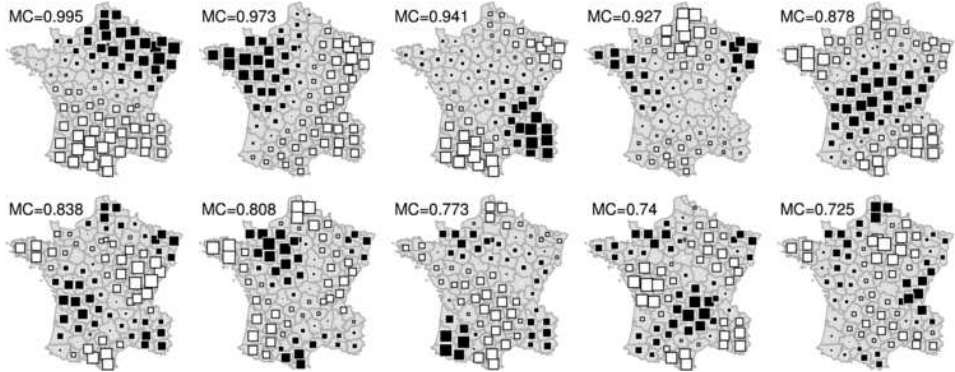}

\caption{Maps of the first ten MEM of the spatial
weighting matrix associated to the neighborhood graph presented on
Figure \protect\ref{fig:fig2}. By definition, MEM are orthogonal vectors
maximizing the values of Moran's coefficient.}\label{fig:fig7}
\end{figure}

We used the spatial weighting matrix associated to the neighborhood
graph presented on Figure \ref{fig:fig2} to construct MEM. The first
ten MEM, corresponding to the highest levels of spatial
autocorrelation, have been mapped in Figure \ref{fig:fig7} and
introduced as spatial explanatory variables in PCAIV. We call this
approach PCAIV-MEM in the rest of the paper. 44.1\% of the total
variance (sum of eigenvalues of PCA) is explained by the first ten MEM
(sum of eigenvalues of PCAIV). The first two dimensions account for
54.9\% and 26.3\%, respectively, of the explained variance. The
outputs of PCAIV-MEM (coefficients of variables, maps of
d\'{e}partement scores, etc.) are not presented, as they are very
similar to those obtained by the previous analyses.\looseness=-1

\subsection{Spatial graph and  weighting matrix}
The MEM framework introduced the spatial information into multivariate
analysis through the eigen-decomposition of the spatial weighting
matrix. Usually, we consider only a~part of the information contained
in this matrix because only a subset of MEM are used as regressors in
PCAIV. In this section we focus on multivariate methods that consider
the spatial weighting matrix under its original form.

\citet{SD381} was the first to introduce a neighborhood graph into
a~multivariate analysis. Following this initial work, many methods have
been mainly developed by the French school of statisticians
[\citet{SD380}; \citet{SD38}; \citet{SD442}]. These contributions were important from a
methodological point of view, but have been rarely used for applied
problems. Indeed, they have a major drawback in their objectives: they
maximize the local variance (i.e., the difference between neighbors),
while users more often want to minimize this quantity and maximize the
spatial correlation (i.e., the $\mathit{SMOOTH}$).

\citet{SD694} was the first to develop a multivariate analysis based on
MC. His work considered only normed and centered variables (i.e.,
normed PCA) for the multivariate part and a binary symmetric
connectivity matrix for the spatial aspect. \citet{SD807} generalized
Wartenberg's method by introducing a row-standardized spatial weighting
matrix in the analysis of a statistical triplet $ ( \mathbf{X},
\mathbf{Q}, \mathbf{D}  )$. Hence, this approach is very general
and allows us to define spatially-constrained versions of various
methods (corresponding to different triplets) such as correspondence
analysis or multiple correspondence analysis.

By extension of the lag vector, a lag matrix
$\mathbf{\tilde{X}}=\mathbf{WX}$ can be defined. The two tables
$\mathbf{\tilde{X}}$ and $\mathbf{X}$ are fully matched, that is, they
have the same columns (variables) and rows (observations). MULTISPATI
(Multivariate spatial analysis based on Moran's index) aims to identify
multivariate spatial structures by studying the link between
$\mathbf{\tilde{X}}$ and $\mathbf{X}$ using the coinertia analysis
[\citet{SD151}; \citet{SD162}] of a~pair of fully matched tables
[\citet{SD667}; \citet{SD161}]. It corresponds to the analysis of the statistical
triplet $ ( \mathbf{X}, \mathbf{Q},
\frac{1}{2}(\mathbf{W}^\mathrm{T}\mathbf{D} + \mathbf{DW}) )$.
The objective of the analysis is to find a~vector \textbf{a} (with
$\Vert\mathbf{a} \Vert^2_\mathbf{Q}$) maximizing the quantity
defined in equation (\ref{eq0a}):
\begin{eqnarray}\label{eq4}
Q(\mathbf{a}) &=&  \mathbf{a}^\mathrm{T}\mathbf{Q}^\mathrm{T}\mathbf{X}^\mathrm{T}\tfrac{1}{2}(\mathbf{W}^\mathrm{T}\mathbf{D}^\mathrm{T} + \mathbf{DW})\mathbf{XQa}\nonumber\\
&=& \tfrac{1}{2} (\mathbf{a}^\mathrm{T}\mathbf{Q}^\mathrm{T}\mathbf{X}^\mathrm{T}\mathbf{W}^\mathrm{T}\mathbf{D}^\mathrm{T} \mathbf{XQa} + \mathbf{a}^\mathrm{T}\mathbf{Q}^\mathrm{T}\mathbf{X}^\mathrm{T}
\mathbf{DWXQa})\nonumber\\[-8pt]\\[-8pt]
&=& \tfrac{1}{2} \langle \mathbf{XQa},\mathbf{WXQa} \rangle_\mathbf{D} + \langle \mathbf{WXQa},\mathbf{XQa} \rangle _\mathbf{D}\nonumber\\
&=& \mathbf{a}^\mathrm{T}\mathbf{Q}^\mathrm{T}\mathbf{X}^\mathrm{T}\mathbf{DWXQa}=\mathbf{r}^\mathrm{T}\mathbf{DWr}=\mathbf{r}^\mathrm{T}\mathbf{D\tilde{r}}.\nonumber
\end{eqnarray}

\begin{figure}[b]
\vspace*{-3pt}
\includegraphics{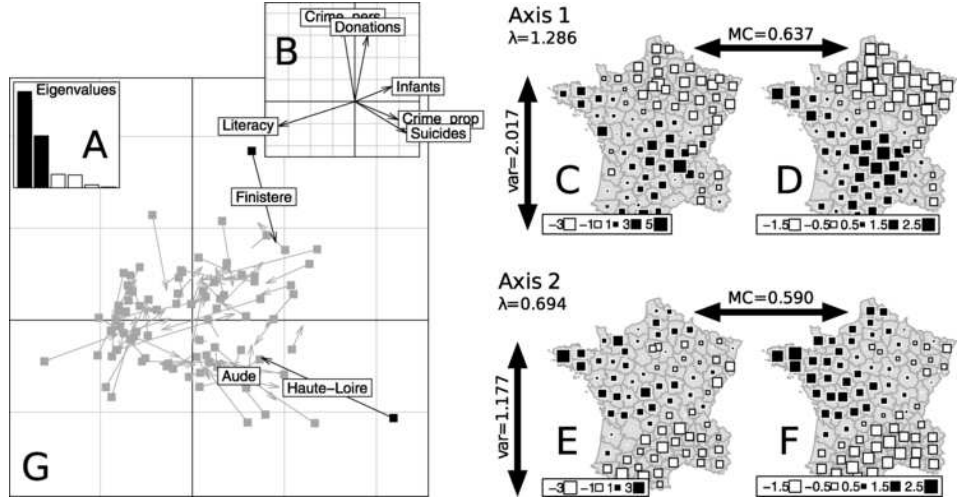}

\caption{$\mathit{MULTISPATI}$ of Guerry's data.
\textup{(A)} Barplot of eigenvalues. \textup{(B)} Coefficients of
variables. Mapping of scores of plots on the first \textup{(C)} and
second \textup{(E)} axis and of lagged scores (averages of neighbors
weighted by the spatial connection matrix) for the first \textup{(D)}
and second \textup{(F)} axis. Representation of scores and lagged
scores \textup{(G)} of plots (for each d\'{e}partement, the arrow links
the score to the lagged score). Only the d\'{e}partements discussed in
the text are indicated by their labels.} \label{fig:fig8}
\end{figure}

This analysis maximizes the scalar product between a linear combination
of original variables ($\mathbf{r}=\mathbf{XQa}$) and a linear
combination of lagged variables ($\mathbf{\tilde{r}}=\mathbf{WXQa}$).
Equation (\ref{eq4}) can be rewritten as
\begin{eqnarray}\label{eq5}
Q(\mathbf{a}) &=&   \frac{\mathbf{a}^\mathrm{T}\mathbf{Q}^\mathrm{T}\mathbf{X}^\mathrm{T} \mathbf{DWXQa}}{\mathbf{a}^\mathrm{T}\mathbf{Q}^\mathrm{T}\mathbf{X}^\mathrm{T} \mathbf{DXQa}}\mathbf{a}^\mathrm{T}\mathbf{Q}^\mathrm{T}\mathbf{X}^\mathrm{T}
\mathbf{DXQa}\nonumber\\[-9pt]\\[-9pt]
&=&   \operatorname{MC}_\mathbf{D}(\mathbf{XQa}) \cdot
\Vert\mathbf{X}\mathbf{Q}\mathbf{a}\Vert^2_\mathbf{D}=\operatorname{MC}_\mathbf{D}(\mathbf{r})
\cdot \Vert\mathbf{r}\Vert^2_\mathbf{D}.\nonumber
\end{eqnarray}
MULTISPATI finds coefficients (\textbf{a}) to obtain a linear
combination of variables ($\mathbf{r}=\mathbf{XQa}$) that maximizes a
compromise between the classical multivariate analysis
($\Vert\mathbf{r}\Vert^2_\mathbf{D}$) and a generalized version of
Moran's coefficient [$\operatorname{MC}_\mathbf{D}(\mathbf{r})$]. The
only difference between the classical Moran's coefficient [equation
(\ref{eq2})] and its generalized version $\operatorname{MC}_\mathbf{D}$
is that the second one used a~general matrix of weights $\mathbf{D}$,
while the first considers only the usual case of uniform weights
($\mathbf{D}=\frac{1}{n} \mathbf{I}_n$).

In practice, the maximum of equation (\ref{eq5}) is obtained for
$\mathbf{a}=\mathbf{a}^1$, where~$\mathbf{a}^1$ is the first
eigenvector of the \textbf{Q}-symmetric matrix
$\frac{1}{2}\mathbf{X}^\mathrm{T}(\mathbf{W}^\mathrm{T}\mathbf{D} +
\mathbf{DW})\mathbf{Q}$. This maximal value is equal to the associated
eigenvalue $\lambda_1$. Further eigenvectors maximize the same quantity
with the additional constraint of orthogonality.

MULTISPATI has been applied to Guerry's data (Figure \ref{fig:fig8}). The
barplot of eigenvalues (Figure \ref{fig:fig8}A) suggests two main spatial
structures. The coefficients used to construct the linear combinations
of variables are represented in Figure~\ref{fig:fig8}B.\vadjust{\goodbreak} The first axis
opposes literacy to property crime, suicides and illegitimate births.
\mbox{The second} axis is aligned mainly with personal crime and donations to
the poor.~The maps of the scores (Figure \ref{fig:fig8}C, E) show that
the spatial structures are very close to those identified by PCA. The
similarity of results between PCA and its spatially optimized version
confirm that the main structures of Guerry's data are spatial.

MULTISPATI maximizes the product between the variance and the spatial
autocorrelation of the scores, while PCA (Figure \ref{fig:fig1})
maximizes only the variance. Hence, there is a loss of variance
compared to PCA (2.14 versus 2.017 for axis 1; 1.201 versus 1.177 for
axis 2) but a gain of spatial autocorrelation (0.551 versus 0.637 for
axis 1; 0.561 versus 0.59 for axis 2).

Spatial autocorrelation can be seen as the link between one variable
and the lagged vector [equation (\ref{eq3})]. This interpretation is used
to construct the Moran scatterplot and can be extended to the
multivariate case in  MULTISPATI by analyzing the link between scores
(Figure \ref{fig:fig8}C, E) and lagged scores (Figure \ref{fig:fig8}D, F).
Each d\'{e}partement can be represented on the factorial map by an
arrow (the bottom corresponds to its score, the head corresponds to its
lagged score, Figure \ref{fig:fig8}G). A short arrow reveals a local
spatial similarity (between one plot and its neighbors), while a long
arrow reveals a spatial discrepancy. This viewpoint can be interpreted
as a multivariate extension of the local index of spatial association
[\citet{SD565}]. For instance, Aude has a very small arrow, indicating
that this d\'{e}partement is very similar to its neighbors. On the
other hand, the arrow for Haute-Loire has a long horizontal length
which reflects its high values for the variables Infants (31017),
Suicides (163241) and Crime\_prop (18043) compared to the average
values over its neighbors (27032.4, 60097.8 and 10540.8 for these three
variables). Finist\`{e}re corresponds to an arrow with a long vertical
length which is due to its high values compared to its neighbors for
Donations (23945 versus 12563) and Crime\_pers (29872 versus 25962).

\section{Conclusions}

We have presented different ways of incorporating the spatial
information in multivariate analysis methods. While PCA is not
constrained, spatial information can be introduced as a partition
(BCA), a~polynomial of geographic coordinates (PCAIV-POLY), a~subset of
Moran's eigenvector maps (PCAIV-MEM) or a spatial neighborhood graph
(MULTISPATI). This variety of constraints induces a diversity of
criteria to be maximized by each method: variance (PCA), variance
explained by a spatial partition (BCA) or by spatial predictors
(PCAIV-POLY, PCAIV-MEM), product of the variance by the spatial
autocorrelation (MULTISPATI). By presenting these methods in the
duality diagram framework, we have shown that these approaches are very
general, and can be applied to virtually any multivariate analysis.

These theoretical considerations have practical implications concerning
the use of these methods in applied studies. PCA is a very
general
method allowing\vadjust{\goodbreak} one to identify the main spatial and nonspatial
structures. BCA maximally separates the groups corresponding to a
spatial partition. It is thus adapted when a study focuses on spatial
structures induced by a partitioning defined a priori (e.g.,
administrative units, etc.). If such an a priori partitioning does not
exist, one can easily define such a partition albeit introducing some
element of subjectivity in the consideration of the spatial
information. This problem is solved by PCAIV-POLY, which uses
polynomials to incorporate the spatial information. Polynomials are
easily constructed, but their use is only satisfactory when the
sampling area is roughly homogeneous and the sampling design is nearly
regular [\citet{SD466}]. Other limitations to their use have been
reported in the literature such as the arbitrary choice of the degree
and their ability to account only for smooth broad-scale spatial
patterns [\citet{SD163}].

The use of graphs and spatial weighting matrices allows the
construction of more efficient and flexible representations of space.
Binary spatial weighting matrices can be constructed using distance
criteria or tools derived from graph theory [\citet{SD333}]; they may
also describe spatial discontinuities, boundaries or physical barriers
in the landscape. Spatial weights can be associated to the binary links
to represent the spatial heterogeneity of the landscape using functions
of geographic distances or least-cost links between sampling locations
[\citet{SD941}] or any other proxies/measures of the potential strength
of connection between the locations. MEM are obtained by the
eigen-decomposition of the spatial weighting matrix \textbf{W}. For a
data set with $n$ observations, this eigen-decomposition produces $n-1$
MEM. Hence, a subset of these spatial predictors must be selected to
avoid overfitting in the multivariate regression step of PCAIV.
Concerning Guerry's data set, we choose the first ten MEM arbitrarily.
Other objective selection procedures have been proposed in the
literature. For instance, the criteria can be based on the minimization
of the autocorrelation in residuals [\citet{SD663}] or on the
maximization of the fit of the model [\citet{SD936}]. Hence, only a~part of the spatial information contained in \textbf{W} (corresponding
to the subset of MEM retained by the selection procedure) is considered
in PCAIV. In MULTISPATI, the spatial weighting matrix is used in its
original form, so that the whole spatial information contained in it is
taken into account in the multivariate analysis.

Even if the methods presented are quite different in their theoretical
and practical viewpoints, their applications to Guerry's data set yield
very similar results. We provided a quantitative measure of this
similarity by computing Procrustes statistics [\citet{SD516};
\citet{SD161}] between the scores of the d\'{e}partements on the first
two axes for the different analyses (Table~\ref{tab:tab3}). All the values of the statistics
are very high and significant; this confirms the high concordance
between the outputs of the different methods. This similarity of
results is due to the very clear\vadjust{\goodbreak} structures of the data set and to the
high level of autocorrelation of these structures (Figure
\ref{fig:fig4}). In this example the main advantage of the
spatially-constrained methods is in the choice of the number of
dimensions to interpret; while the barplot of eigenvalues of PCA can be
difficult to interpret (see above and Figure \ref{fig:fig1}A), it is
clear that two spatial dimensions must be interpreted in BCA (Figure
\ref{fig:fig5}A) or MULTISPATI (Figure \ref{fig:fig8}A).

\begin{table}
\tablewidth=280pt
\caption{Procrustes statistics measuring the concordance between the
scores of the d\'{e}partements on the first two axes of the different
analyses. A~value of 1 indicates a perfect match between two
configurations of d\'{e}partement scores. Randomization procedures with
999 permutations have been used to test the significance of the
concordance. All the statistics are significant ($p=0.001$)}\label{tab:tab3}
\begin{tabular*}{280pt}{@{\extracolsep{\fill}}lcccc@{}}
  \hline
 & \textbf{PCA} & \textbf{BCA} & \textbf{PCAIV-POLY} & \textbf{PCAIV-MEM} \\
  \hline
BCA & 0.979 &  &  &  \\
  PCAIV-POLY & 0.979 & 0.990 &  &  \\
  PCAIV-MEM & 0.989 & 0.994 & 0.995 &  \\
  MULTISPATI & 0.987 & 0.995 & 0.995 & 0.999 \\
   \hline
\end{tabular*}
\vspace*{-3pt}
\end{table}

In the case of Guerry's data, the very simple and clear-cut structures
seem to be recovered by all the approaches presented here. In more
complex data sets, spatially constrained methods prove superior to
standard approaches for detecting spatial multivariate patterns.
\citet{SD807} presented an example where a standard multivariate method
was unable to identify any structure and is outperformed by MULTISPATI,
which allows us to discover interesting spatial patterns. In general,
if the objective of a~study is to detect spatial patterns, it would be
preferable to use a spatially-constrained method. PCA could also be
useful, but it is designed to identify the main structures that can or
cannot be spatialized. On the other hand, spatial multivariate methods
are optimized to focus on the spatial aspect and would generally
produce clearer and smoother results. The outputs and interpretation
tools of these methods are also more adapted to visualizing and
quantifying the main multivariate spatial structures.

From a methodological viewpoint, these approaches provide new ways of
taking into account the complexity of sampling designs in the framework
of multivariate methods. Following the famous paper of \citet{SD399},
the analysis of spatial structures has been a major issue in community
ecology and originated several methodological developments in the field
of spatial multivariate analysis. To date, the most integrated and
flexible approaches have used a spatial weighting matrix which can be
seen as a general way to consider spatial proximities. Potential
methodological perspectives are important, as these approaches could
easily be extended to any other sampling constraints that can be
expressed in the form of a matrix of similarities between the
observations.\vadjust{\goodbreak}

\section*{Acknowledgments}
We would like to warmly thank Michael Friendly for freely distributing
Guerry's data set and for providing constructive comments on an earlier
version of the manuscript. We thank Susan Holmes for her invitation to
participate in this special issue.

\begin{supplement}\label{suppA}
\stitle{Implementation in R}
\slink[doi]{10.1214/10-AOAS356SUPP}
\slink[url]{http://lib.stat.cmu.edu/aoas/356/supplement.zip}
\sdatatype{.zip}
\sdescription{This website hosts an R package (\texttt{Guerry})
containing the Guerry's data set (maps and data). The package contains
also a tutorial (vignette) showing how to reproduce the analyses and
the graphics presented in this paper using mainly the  package
\texttt{ade4} [\citet{SD839}]. The package \texttt{Guerry} is also
available on CRAN and can be installed using the
\texttt{install.packages(``Guerry'')} command in a R
session.}
\end{supplement}

\printaddresses

\end{document}